\documentclass[11pt]{article}%
\usepackage{amsmath}
\usepackage{amsfonts}
\usepackage{amssymb}
\usepackage{graphicx}
\usepackage{fullpage}%
\providecommand{\U}[1]{\protect\rule{.1in}{.1in}}
\newtheorem{theorem}{Theorem}

\newtheorem{lemma}[theorem]{Lemma}

\newtheorem{proposition}[theorem]{Proposition}

\newenvironment{proof}[1][Proof]{\noindent\textbf{#1.} }{\ \rule{0.5em}{0.5em}}
\begin{document}

\title{Closed Timelike Curves Make Quantum\\and Classical Computing Equivalent }
\author{Scott Aaronson\thanks{Email: aaronson@csail.mit.edu.}\\MIT
\and John Watrous\thanks{Email: watrous@cs.uwaterloo.ca.}\\University of Waterloo}
\date{}
\maketitle

\begin{abstract}
While closed timelike curves (CTCs) are not known to exist, studying their
consequences has led to nontrivial insights in general relativity, quantum
information, and other areas. \ In this paper we show that if CTCs existed,
then quantum computers would be no more powerful than classical computers:
both would have the (extremely large) power of the complexity class
$\mathsf{PSPACE}$, consisting of all problems solvable by a conventional
computer using a polynomial amount of memory. \ This solves an open problem
proposed by one of us in 2005, and gives an essentially complete understanding
of computational complexity in the presence of CTCs. \ Following the work of
Deutsch, we treat a CTC as simply a region of spacetime where a
\textquotedblleft causal consistency\textquotedblright\ condition is imposed,
meaning that Nature has to produce a (probabilistic or quantum) fixed-point of
some evolution operator. \ Our conclusion is then a consequence of the
following theorem: given any quantum circuit (not necessarily unitary), a
fixed-point of the circuit can be (implicitly) computed in polynomial space.
\ This theorem might have independent applications in quantum information.

\end{abstract}

\section{Introduction\label{INTRO}}

The possibility of closed timelike curves (CTCs) within general relativity and
quantum gravity theories has been studied for almost a
century\ \cite{godel:ctc,stockum,mty}. \ A different line of research has
sought to understand the \textit{implications} of CTCs, supposing they
existed, for quantum mechanics, computation, and information
\cite{deutsch:ctc,brun,bacon}.

In this paper we contribute to the latter topic, by giving the first complete
characterization of the computational power of CTCs. \ We show that if CTCs
existed, then both classical and quantum computers would have exactly the
power of the complexity class $\mathsf{PSPACE}$, which consists of all
problems solvable on a classical computer with a polynomial amount of memory.
\ To put it differently, CTCs would make polynomial time equivalent to
polynomial space as computational resources, and would also make quantum and
classical computers equivalent to each other in their computational power.
\ Our results treat CTCs using the \textquotedblleft causal
consistency\textquotedblright\ framework of Deutsch \cite{deutsch:ctc}.

It will not be hard to show that classical computers with CTCs can simulate
$\mathsf{PSPACE}$ and be simulated in it (though as far as we know, this
result is new). \ The main difficulty will be to show that quantum computers
with CTCs can be simulated in $\mathsf{PSPACE}$. \ To prove this, we need to
give an algorithm for (implicitly) computing fixed-points of superoperators in
polynomial space. \ Our algorithm relies on fast parallel algorithms for
linear algebra due to Borodin, Cook, and Pippenger \cite{bcp}, and might be of
independent interest.

The paper is organized as follows. \ In Section \ref{BACKGROUND}, we explain
needed background about Deutsch's causal consistency framework and
computational complexity, and review previous work by Bacon \cite{bacon}, Brun
\cite{brun},\ and Aaronson \cite{aar:np}. \ In Section \ref{CLASSICAL}, we
show that classical computers with CTCs have exactly the power of
$\mathsf{PSPACE}$. \ Section \ref{QUANTUM} extends the analysis of Section
\ref{CLASSICAL}\ to show that \textit{quantum} computers with CTCs have
exactly the power of $\mathsf{PSPACE}$. \ In that section, we make the
simplifying assumption that all quantum gates can be applied perfectly and
that amplitudes are rational. \ In Section \ref{ERROR}, we consider what
happens when gates are subject to finite error, and extend previous work of
Bacon \cite{bacon}\ to show that quantum computers with CTCs can solve
$\mathsf{PSPACE}$\ problems in a \textquotedblleft
fault-tolerant\textquotedblright\ way. \ We conclude in Section \ref{DISC}
with some general remarks and open problems.

\section{Background\label{BACKGROUND}}

\subsection{Causal Consistency\label{CONSIS}}

It was once believed that CTCs would lead inevitably to logical
inconsistencies such as the Grandfather Paradox. \ But in a groundbreaking
1991 paper, Deutsch \cite{deutsch:ctc}\ showed that this intuition fails,
provided the physics of the CTC is quantum-mechanical. \ Deutsch's insight was
that a CTC should simply be regarded as a region of spacetime where Nature
enforces a requirement of \textit{causal consistency}: in other words, that
the evolution operator within that region should map the state of the initial
hypersurface to itself. \ Given the evolution operator $f$, Nature's
\textquotedblleft task\textquotedblright\ is thus to find a fixed-point of
$f$: that is, an input $x$ such that $f\left(  x\right)  =x$. \ Of course, not
every deterministic evolution operator $f$ \textit{has} a fixed-point: that is
just one way of stating the Grandfather Paradox. \ On the other hand, it is a
basic linear-algebra fact that every quantum operation $\Phi$ has a
fixed-point: that is, a density matrix $\rho$\ such that $\Phi\left(
\rho\right)  =\rho$. \ For any $\Phi$, such a $\rho$ can then be used to
produce a CTC evolution that satisfies the causal consistency requirement.
\ So for example, a consistent resolution of the Grandfather Paradox is that
you are born with $1/2$\ probability, and \textit{if} you are born you go back
in time to kill your grandfather, therefore you are born with $1/2$
probability, etc.

Notice that Deutsch's resolution works just as well in classical probabilistic
theories as in quantum-mechanical ones. \ For just as every quantum operation
has a fixed-point, so every Markov chain has a stationary distribution. \ What
matters is simply that the state space and the set of transformations are such
that fixed-points exist.

Although CTCs need not lead to inconsistencies, Deutsch pointed out that they
\textit{would} have striking consequences for the theory of computing. \ As an
example, CTCs could be exploited to solve $\mathsf{NP}$-complete and other
\textquotedblleft intractable\textquotedblright\ computational problems using
only polynomial resources. \ To see this, suppose some integers $x\in\left\{
0,1,\ldots,2^{n-1}\right\}  $\ are \textquotedblleft
solutions\textquotedblright\ and others are not, and that our goal is to find
a solution in time polynomial in $n$, assuming solutions exist and can be
recognized efficiently. \ Then we could build a machine $M$\ that applied the
following transformation to its input $x$: if $x$ is a solution
then\ $M\left(  x\right)  =x$, while if $x$ is not a solution then $M\left(
x\right)  =\left(  x+1\right)  \operatorname{mod}2^{n}$. \ Now suppose we use
a CTC to feed $M$ its own output as input. \ Then it is not hard to see that
the only way for the evolution to satisfy causal consistency is for $M$ to
input, and output, a solution.

In this way, an exponentially-hard computational problem could get solved
without exponential effort ever being invested to solve it, merely because
that is the only way to satisfy causal consistency. \ A rough analogy would be
Shakespeare's plays being written by someone from the present going back in
time and dictating the plays to him.

It is sometimes said that if CTCs existed, then one could \textit{obviously}
do computations of unlimited length in an instant, by simply computing the
answer, then sending it back in time to before one started.\ \ However, this
proposal does not work for two reasons. \ First, it ignores the Grandfather
Paradox: what happens if, on receiving the output, one goes back in time and
changes the input? \ Second,\ it is perhaps unclear why a computation lasting
$10^{1000}$ years should be considered \textquotedblleft
feasible,\textquotedblright\ merely because we are able to obtain the solution
\textit{before} performing the computation. \ It seems that an honest
accounting should require the computations performed inside the CTC to be
efficient (say, polynomial-time), with any computational speedup coming from
the requirement of causal consistency.

\subsection{Complexity Theory\label{COMPLEXITY}}

For background on classical computational complexity theory, see for example
Arora and Barak \cite{arorabarak}; for a recent survey of quantum complexity
theory, see Watrous \cite{watrous:survey}. \ Here, we briefly describe the
main complexity classes we will consider. $\mathsf{PSPACE}$ (Polynomial Space)
is the class of decision problems that are solvable by a classical computer,
using an amount of memory that is bounded by a polynomial function of the size
of the input $n$ (but possibly an exponential amount of time). \ An example of
such a problem is, given a configuration of an $n\times n$\ Go board, to
decide whether White has a winning strategy using $n^{2}$\ or fewer moves.
\ $\mathsf{NP}$\ (Nondeterministic Polynomial-Time) is the class of decision
problems for which every \textquotedblleft yes\textquotedblright\ answer has a
polynomial-time-checkable, polynomial-size proof or witness. \ $\mathsf{NP}%
$-complete problems are, loosely speaking, the \textquotedblleft
hardest\textquotedblright\ problems in $\mathsf{NP}$: that is, those
$\mathsf{NP}$\ problems to which all other $\mathsf{NP}$\ problems can be
efficiently reduced. \ An example is, given a graph, to decide whether it has
a Hamiltonian cycle (that is, a cycle that visits each vertex exactly once).
\ $\mathsf{PSPACE}$ contains $\mathsf{NP}$ (thus, in particular, the
$\mathsf{NP}$-complete problems)---since in polynomial space, one can simply
loop over all possible witnesses and see if any of them are correct.
\ However, $\mathsf{PSPACE}$\ is believed to be considerably larger than
$\mathsf{NP}$. \ So in saying that computers with CTCs can efficiently solve
$\mathsf{PSPACE}$\ problems, we are saying something stronger than just that
they can solve $\mathsf{NP}$-complete problems.

Our main result is that computers with CTCs have precisely the power of
$\mathsf{PSPACE}$, and that this is true whether the computers are classical
or quantum. \ Previously, Watrous \cite{watrous:space} showed that
$\mathsf{BQPSPACE}$ (Bounded-Error Quantum Polynomial Space) is equal to
$\mathsf{PSPACE}$: that is, any problem solvable by a quantum computer with
polynomial memory is also solvable by a classical computer with polynomial
memory. \ (By contrast, quantum computers are conjectured to offer an
exponential improvement over classical computers in \textit{time}.) \ Here, we
show that quantum computers are polynomially equivalent to classical computers
in the CTC setting as well.

\subsection{Related Work\label{RELATED}}

Besides Deutsch's paper \cite{deutsch:ctc}, we know of three other works
directly relevant to computational complexity in the presence of CTCs.
\ First, Bacon \cite{bacon} showed that $\mathsf{NP}$-complete problems can be
solved with polynomial resources,\ even using CTCs that are only
\textquotedblleft one bit wide\textquotedblright\ (i.e., able to transmit a
single qubit or probabilistic classical bit back in time).\footnote{On the
other hand, Bacon's approach would require a polynomial number of such CTC's,
rather than a single CTC as in Deutsch's approach.} \ Bacon also showed that,
using his approach, one can solve not only $\mathsf{NP}$ problems but even
$\mathsf{\#P}$ problems, which involve \textit{counting} solutions rather than
just finding one. \ (The class $\mathsf{\#P}$---or more formally its decision
version $\mathsf{P}^{\mathsf{\#P}}$---is a subclass of $\mathsf{PSPACE}$, with
the containment believed to be strict.) \ Finally, Bacon showed that
techniques from the theory of quantum fault-tolerance could be used to make
certain CTC computations, including the ones used to solve $\mathsf{\#P}$
problems, robust to small errors.

Second, Brun \cite{brun} claimed to show that CTCs would allow the efficient
solution of any problem in $\mathsf{PSPACE}$. \ However, Brun did not specify
the model of computation underlying his results, and the most natural
interpretation of his \textquotedblleft CTC algorithms\textquotedblright%
\ would appear to preclude their solving $\mathsf{PSPACE}$\ problems. \ For
Brun, a fixed-point of a CTC evolution seems to be necessarily
\textit{deterministic}---in which case, finding such a fixed-point is an
$\mathsf{NP}$\ problem (note that $\mathsf{NP}$\ is almost universally
believed to be smaller than $\mathsf{PSPACE}$). \ Thus, to prove that
classical \textit{or} quantum computers with CTCs give the full power of
$\mathsf{PSPACE}$, it seems essential to adopt Deutsch's causal consistency
model (which Brun does not discuss).

Third, as part of a survey on \textquotedblleft$\mathsf{NP}$-complete Problems
and Physical Reality\textquotedblright\ \cite{aar:np}, Aaronson sketched the
definitions of $\mathsf{P}_{\mathsf{CTC}}$\ and $\mathsf{BQP}_{\mathsf{CTC}}%
$\ (classical and quantum polynomial time with CTCs) that we adopt in this
paper. \ He also sketched a proof that $\mathsf{PSPACE}=\mathsf{P}%
_{\mathsf{CTC}}\subseteq\mathsf{BQP}_{\mathsf{CTC}}\subseteq\mathsf{EXP}$.
\ That is, classical computers with CTCs have exactly the power of polynomial
space, while quantum computers with CTCs have at least the power of polynomial
space and at most the power of classical exponential time. \ The key problem
that Aaronson left open was to pin down the power of quantum computers with
CTCs precisely. \ This is the problem we solve in this paper.

\section{The Classical Case\label{CLASSICAL}}

To state our results, it is crucial to have a formal model of computation in
the presence of CTCs.

We define a \textit{deterministic CTC algorithm} $\mathcal{A}$ to be
deterministic polynomial-time algorithm that takes as input a string
$x\in\left\{  0,1\right\}  ^{n}$, and that produces as output a Boolean
circuit $C=C_{x}$, consisting of AND, OR, and NOT gates. \ The circuit $C$
acts on bits in two registers: a \textit{CTC register} $\mathcal{R}_{CTC}$,
and a \textit{causality-respecting register} $\mathcal{R}_{CR}$. \ The
registers $\mathcal{R}_{CTC}$\ and $\mathcal{R}_{CR}$\ consist of $p\left(
n\right)  $\ and $q\left(  n\right)  $\ bits respectively, for some
polynomials $p$ and $q$ depending on $\mathcal{A}$. \ Thus, $C$ can be seen as
a Boolean function $C:\left\{  0,1\right\}  ^{p\left(  n\right)  +q\left(
n\right)  }\rightarrow\left\{  0,1\right\}  ^{p\left(  n\right)  +q\left(
n\right)  }$, which maps an ordered pair $\left\langle y,z\right\rangle
\in\mathcal{R}_{CTC}\times\mathcal{R}_{CR}$ to another ordered pair $C\left(
\left\langle y,z\right\rangle \right)  $.

For convenience, we assume that the causality-respecting register
$\mathcal{R}_{CR}$\ is initialized to $0^{q\left(  n\right)  }$. \ The CTC
register, on the other hand, must be initialized to some probability
distribution over $p\left(  n\right)  $-bit strings that will ensure causal
consistency. \ More formally, let $\mathcal{D}$ be a probability distribution
over $\mathcal{R}_{CTC}\times\mathcal{R}_{CR}$,\ and let $C\left(
\mathcal{D}\right)  $\ be the distribution over $\mathcal{R}_{CTC}%
\times\mathcal{R}_{CR}$ induced by drawing a sample from $\mathcal{D}$\ and
then applying $C$ to it. \ Also, let $\left[  \cdot\right]  _{CTC}$\ be an
operation that discards the causality-respecting register (i.e., marginalizes
it out), leaving only the CTC register. \ Then we need the initial probability
distribution $\mathcal{D}$\ over $\mathcal{R}_{CTC}\times\mathcal{R}_{CR}$\ to
satisfy the following two conditions:

\begin{enumerate}
\item[(i)] $\mathcal{D}$ has support only on pairs of the form $\left\langle
y,0^{q\left(  n\right)  }\right\rangle $.

\item[(ii)] $\mathcal{D}$ satisfies the \textit{causal consistency equation}
$\left[  \mathcal{D}\right]  _{CTC}=\left[  C\left(  \mathcal{D}\right)
\right]  _{CTC}$.
\end{enumerate}

We claim that such a $\mathcal{D}$\ always exists. \ This is easy to prove:
let $C^{\prime}\left(  y\right)  :=\left[  C\left(  \left\langle y,0^{q\left(
n\right)  }\right\rangle \right)  \right]  _{CTC}$ be the induced circuit that
acts only on the CTC register. \ Then it suffices to find a distribution
$\mathcal{D}^{\prime}$\ over $\mathcal{R}_{CTC}$\ such that $C^{\prime}\left(
\mathcal{D}^{\prime}\right)  =\mathcal{D}^{\prime}$. \ To find such a
$\mathcal{D}^{\prime}$, we consider the graph of the function $C^{\prime
}:\left\{  0,1\right\}  ^{p\left(  n\right)  }\rightarrow\left\{  0,1\right\}
^{p\left(  n\right)  }$, find a cycle in that graph (which must exist, since
the graph is finite), and let $\mathcal{D}^{\prime}$\ be the uniform
distribution over points in the cycle. \ Finally we set $\mathcal{D}%
=\left\langle \mathcal{D}^{\prime},0^{q\left(  n\right)  }\right\rangle $.%
\begin{figure}
[ptb]
\begin{center}
\includegraphics[
trim=2.087492in 3.915504in 2.090727in 0.000000in,
height=1.6259in,
width=2.565in
]%
{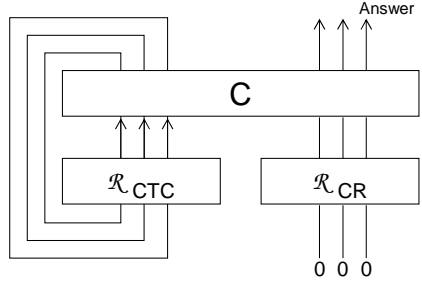}%
\caption{Diagram of a classical CTC computer. \ A circuit $C$\ performs a
polynomial-time computation involving \textquotedblleft closed timelike curve
bits\textquotedblright\ (the register $\mathcal{R}_{CTC}$) as well as
\textquotedblleft causality-respecting bits\textquotedblright\ (the register
$\mathcal{R}_{CR}$). \ Nature must then find a probability distribution over
$\mathcal{R}_{CTC}$\ that satisfies Deutsch's causal consistency equation.
\ The final answer is read out from $\mathcal{R}_{CR}$.}%
\label{ctcfig}%
\end{center}
\end{figure}

We are now ready to define the complexity class $\mathsf{P}_{\mathsf{CTC}}$,
of problems solvable using classical computers with CTCs. \ We say that a CTC
algorithm $\mathcal{A}$\ \textit{accepts} the input $x$ if, for every
distribution $\mathcal{D}$\ satisfying conditions (i) and (ii) above,
$C\left(  \mathcal{D}\right)  $\ has support only on pairs of the form
$\left\langle y,z1\right\rangle $ (i.e., such that the last bit of the
causality-respecting register is a $1$). \ Likewise, we say $\mathcal{A}%
$\ \textit{rejects} $x$ if for every $\mathcal{D}$\ satisfying (i) and (ii),
$C\left(  \mathcal{D}\right)  $\ has support only on pairs of the form
$\left\langle y,z0\right\rangle $. \ (Of course, it is possible that $C\left(
\mathcal{D}\right)  $\ has support on both kinds of pairs, in which case
$\mathcal{A}$\ neither accepts nor rejects.) \ We say $\mathcal{A}%
$\ \textit{decides the language} $L\subseteq\left\{  0,1\right\}  ^{\ast}$\ if
$\mathcal{A}$\ accepts every input $x\in L$, and rejects every input $x\notin
L$. \ Then $\mathsf{P}_{\mathsf{CTC}}$\ is the class of all languages
$L$\ that are decided by some deterministic CTC algorithm.

Let us make a few remarks about the definition. \ First, the requirement that
some polynomial-time algorithm $\mathcal{A}$\ output the circuit $C=C_{x}$\ is
intended to prevent hard-to-compute information\ from being hard-wired into
the circuit. \ This requirement is standard in complexity theory; it is also
used, for example, in the definition of $\mathsf{BQP}$. \ Second, our
definition required $C$ to succeed with certainty, and did not allow $C$ to
introduce its own randomness, besides that produced by the causal consistency
condition. \ We could relax these requirements to obtain the complexity class
$\mathsf{BPP}_{\mathsf{CTC}}$, or bounded-error probabilistic polynomial time
with access to a CTC. \ However, it will turn out that $\mathsf{P}%
_{\mathsf{CTC}}=\mathsf{BPP}_{\mathsf{CTC}}=\mathsf{PSPACE}$ anyway.

\subsection{Results\label{CRESULTS}}

We now prove $\mathsf{P}_{\mathsf{CTC}}=\mathsf{PSPACE}$.

\begin{lemma}
\label{inpspace}$\mathsf{P}_{\mathsf{CTC}}\subseteq\mathsf{PSPACE}$.
\end{lemma}

\begin{proof}
Let $C$\ be a polynomial-size circuit that maps $\mathcal{R}_{CTC}%
\times\mathcal{R}_{CR}$\ to itself, as in the definition of $\mathsf{P}%
_{\mathsf{CTC}}$. \ Then our $\mathsf{PSPACE}$\ simulation algorithm is as
follows. \ First, let $C^{\prime}\left(  y\right)  :=\left[  C\left(
\left\langle y,0^{q\left(  n\right)  }\right\rangle \right)  \right]  _{CTC}$
be the induced circuit that acts only on $\mathcal{R}_{CTC}$. \ Then given a
string $y\in\left\{  0,1\right\}  ^{p\left(  n\right)  }$, say $y$ is
\textit{cyclic} if $C^{\prime\left(  k\right)  }\left(  y\right)  =y$\ for
some positive integer $k$. \ In other words, $y$ is cyclic if repeated
application of $C^{\prime}$ takes us from $y$ back to itself. \ Clearly every
$C^{\prime}$ has at least one cyclic string. \ Furthermore, it is clear from
the definition of $\mathsf{P}_{\mathsf{CTC}}$\ that if $x\in L$\ then every
cyclic string must lead to an output of $1$ in the last bit of $\mathcal{R}%
_{CR}$, while if $x\notin L$\ then every cyclic string must lead to an output
of $0$. \ Hence the problem essentially reduces to finding a cyclic string.

But it is easy to find a cyclic string in polynomial space: the string
$y^{\ast}:=C^{\prime(2^{p\left(  n\right)  })}\left(  y\right)  $ will be
cyclic for any $y$. \ The one remaining step is to compute $C\left(
\left\langle y^{\ast},0^{q\left(  n\right)  }\right\rangle \right)  $, and
then output the last bit of $\mathcal{R}_{CR}$.
\end{proof}

We are indebted to Lance Fortnow for the following lemma.

\begin{lemma}
\label{inpctc}$\mathsf{PSPACE}\subseteq\mathsf{P}_{\mathsf{CTC}}$.
\end{lemma}

\begin{proof}
For some polynomial $p$, let $M$ be a $p\left(  n\right)  $-space Turing
machine (i.e. every configuration of $M$ takes $p\left(  n\right)  $\ bits to
describe). \ We can assume without loss of generality that $M$ includes a
\textquotedblleft clock,\textquotedblright\ which is incremented at every time
step, and which causes $M$ to accept automatically once it reaches its maximum
value. \ This prevents $M$ from ever going into an infinite loop, regardless
of its starting configuration.

Let $m_{1},\ldots,m_{T}$ be the successive configurations of $M$ when run on
an input $x\in\left\{  0,1\right\}  ^{n}$. \ Then our task is to decide, using
a CTC computer, whether $m_{T}$\ is an accepting or a rejecting configuration.

Our CTC algorithm $\mathcal{A}$\ will produce a circuit $C$ that acts on two
registers: a $\left(  p\left(  n\right)  +1\right)  $-bit CTC register
$\mathcal{R}_{CTC}$, and a one-bit causality-respecting register
$\mathcal{R}_{CR}$. \ For simplicity, we start by describing the induced
circuit $C^{\prime}$\ that acts on $\mathcal{R}_{CTC}$. \ Given a
configuration $m$ of $M$, let $S\left(  m\right)  $\ be the successor of $m$:
that is, the configuration obtained from $m$ by incrementing the clock and
performing one step of computation. \ Then the circuit $C^{\prime}$\thinspace
acts as follows, on ordered pairs $\left\langle m,b\right\rangle $ consisting
of a configuration $m$ and a \textquotedblleft control bit\textquotedblright%
\ $b$:

\begin{itemize}
\item If $m$ is neither an accepting nor a rejecting configuration, then
$C^{\prime}\left(  \left\langle m,b\right\rangle \right)  =\left\langle
S\left(  m\right)  ,b\right\rangle $.

\item If $m$ is an accepting configuration, then $C^{\prime}\left(
\left\langle m,b\right\rangle \right)  =\left\langle m_{1},1\right\rangle $.

\item If $m$ is a rejecting configuration, then $C^{\prime}\left(
\left\langle m,b\right\rangle \right)  =\left\langle m_{1},0\right\rangle $.
\end{itemize}

In other words, if $m$ produces an output then $C^{\prime}$ sets the control
bit to that output and goes back to the starting configuration; otherwise
$C^{\prime}$ increments the computation and leaves the control bit unchanged
(see Figure \ref{pspacefig}).%
\begin{figure}
[ptb]
\begin{center}
\includegraphics[
trim=2.609365in 3.392844in 2.612600in 0.000000in,
height=1.8273in,
width=2.1638in
]%
{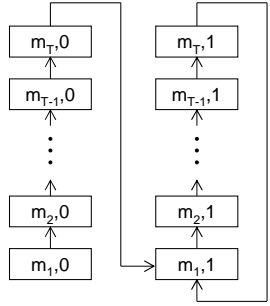}%
\caption{To simulate a $\mathsf{PSPACE}$ machine with a CTC, we perform a
computation for which the only causally consistent evolution is a loop over
all configurations of the machine, with a control bit $b$\ set to its final
value\ (in this example $b=1$).}%
\label{pspacefig}%
\end{center}
\end{figure}

Now consider the graph of the function $C^{\prime}:\left\{  0,1\right\}
^{p\left(  n\right)  +1}\rightarrow\left\{  0,1\right\}  ^{p\left(  n\right)
+1}$. \ It is not hard to see that the only cycle in this graph is $\left(
\left\langle m_{1},1\right\rangle ,\ldots,\left\langle m_{T},1\right\rangle
\right)  $\ if $m_{T}$\ accepts, or $\left(  \left\langle m_{1},0\right\rangle
,\ldots,\left\langle m_{T},0\right\rangle \right)  $\ if $m_{T}$ rejects.
\ Indeed, this is true even if there are configurations that are not reachable
from $\left\langle m_{1},1\right\rangle $\ or $\left\langle m_{1}%
,0\right\rangle $, since those configurations will ultimately lead back to
either $\left\langle m_{1},1\right\rangle $\ or $\left\langle m_{1}%
,0\right\rangle $ and therefore not produce new cycles. \ In other words, the
only cycle is a loop over $m_{1},\ldots,m_{T}$, with the control bit $b$ set
to $M$'s final output. \ Therefore, the only probability distribution
$\mathcal{D}^{\prime}$\ over $\left\{  0,1\right\}  ^{p\left(  n\right)  +1}%
$\ that is \textit{stationary}, in the sense that $C^{\prime}\left(
\mathcal{D}^{\prime}\right)  =\mathcal{D}^{\prime}$, is the uniform
distribution over $\left\langle m_{1},b\right\rangle ,\ldots,\left\langle
m_{T},b\right\rangle $\ where $b$ is $M$'s final output.

Finally, the full circuit $C$\ simply applies $C^{\prime}$\ to $\mathcal{R}%
_{CTC}$, and then copies the control bit into the causality-respecting register.
\end{proof}

\section{The Quantum Case\label{QUANTUM}}

Let $G$ be a universal set of quantum gates, with amplitudes having rational
real and imaginary parts. Then a \textit{quantum CTC algorithm} $\mathcal{A}$
is a deterministic polynomial-time algorithm that takes as input a string
$x\in\left\{  0,1\right\}  ^{n}$, and that produces as output an encoding of a
unitary quantum circuit $Q=Q_{x}$ with gates from $G$.

The circuit $Q$ acts on two quantum registers: a $q\left(  n\right)  $-qubit
CTC register $\mathcal{R}_{CTC}$ and an $r\left(  n\right)  $-qubit
causality-respecting register $\mathcal{R}_{CR}$. The causality-respecting
register $\mathcal{R}_{CR}$ is initialized to $\left\vert 0\right\rangle
^{\otimes r\left(  n\right)  }$, while the CTC register must be initialized to
some $q\left(  n\right)  $-qubit mixed state $\rho$ that will ensure causal
consistency. More formally, we require that
\begin{equation}
\operatorname*{Tr}\nolimits_{\mathcal{R}_{CR}}\left(  Q\left(  \rho
\otimes\left(  |0\rangle\langle0|\right)  ^{\otimes r(n)}\right)  Q^{\dagger
}\right)  =\rho, \label{eq:fixed}%
\end{equation}
which is equivalent to $\rho$ being a fixed-point of the quantum operation
defined as
\[
\Phi\left(  \rho\right)  :=\operatorname*{Tr}\nolimits_{\mathcal{R}_{CR}%
}\left(  Q\left(  \rho\otimes\left(  |0\rangle\langle0|\right)  ^{\otimes
r(n)}\right)  Q^{\dagger}\right)  .
\]
Deutsch \cite{deutsch:ctc} proved that every such quantum operation
\textit{has} a fixed-point, and an alternate proof of this fact follows from
our results in Section~\ref{FIXEDPOINT} below.

We can now define the complexity class $\mathsf{BQP}_{\mathsf{CTC}}$, of
problems solvable using quantum computers with CTCs. Let $\mathcal{M}$ be a
measurement of the last qubit of $\mathcal{R}_{CR}$ in the computational
basis. Then we say the algorithm $\mathcal{A}$ \textit{accepts} $x$ if for
every mixed state $\rho$ satisfying equation \eqref{eq:fixed} above,
$\mathcal{M}\left(  Q\left(  \rho\otimes\left(  |0\rangle\langle0|\right)
^{\otimes r(n)}\right)  Q^{\dagger}\right)  $ results in output $1$ with
probability at least $2/3$. We say $\mathcal{A}$ \textit{rejects} $x$ if for
every $\rho$ satisfying the equation, $\mathcal{M}\left(  Q\left(  \rho
\otimes\left(  |0\rangle\langle0|\right)  ^{\otimes r(n)}\right)  Q^{\dagger
}\right)  $ results in output $1$ with probability at most $1/3$. \ We say
$\mathcal{A}$ decides the language $L\subseteq\left\{  0,1\right\}  ^{\ast}%
$\ if $\mathcal{A}$\ accepts every input $x\in L$, and rejects every input
$x\notin L$. Then $\mathsf{BQP}_{\mathsf{CTC}}$\ is the class of all languages
$L$ that are decided by some quantum CTC algorithm.

In what follows, we develop some needed background, and then prove the main
result that $\mathsf{BQP}_{\mathsf{CTC}}\subseteq\mathsf{PSPACE}$.

\subsection{Matrix Representation of Superoperators\label{MATREP}}

We will make use of a simple way of representing quantum operations as
matrices. \ This representation begins with a representation of density
matrices as vectors by the linear function defined on standard basis states
as
\[
\operatorname*{vec}\left(  \left\vert x\right\rangle \left\langle y\right\vert
\right)  =\left\vert x\right\rangle \left\vert y\right\rangle .
\]
If $\rho$ is an $N\times N$ density matrix, then $\operatorname*{vec}(\rho)$
is the $N^{2}$-dimensional column vector obtained by stacking the rows of
$\rho$ on top of one another. For example,
\[
\operatorname*{vec}%
\begin{pmatrix}
\alpha & \beta\\
\gamma & \delta
\end{pmatrix}
=%
\begin{pmatrix}
\alpha\\
\beta\\
\gamma\\
\delta
\end{pmatrix}
.
\]

Now, suppose that $\Phi$ is a given quantum operation acting on an $N$
dimensional system, meaning that $\Phi:\mathbb{C}^{N\times N}\rightarrow
\mathbb{C}^{N\times N}$\ is linear, completely positive, and trace-preserving.
Given that the effect of $\Phi$ on density matrices is linear, there must
exist an $N^{2}\times N^{2}$ matrix $K(\Phi)$ that satisfies
\[
K(\Phi)\operatorname*{vec}(\rho)=\operatorname*{vec}\left(  \Phi(\rho)\right)
\]
for every possible $N\times N$ density matrix $\rho$. The matrix $K(\Phi)$ is
called the \textit{natural matrix representation} of the quantum operation
$\Phi$, and is uniquely determined by $\Phi$.

The natural matrix representation can easily be calculated from other standard
forms. For example, if an operation $\Phi$ is represented in the usual Kraus
form as
\[
\Phi(\rho)=\sum_{j=1}^{k}A_{j}\rho A_{j}^{\dagger},
\]
then it holds that
\[
\operatorname{vec}\left(  \Phi\left(  \rho\right)  \right)  =\left(
\sum_{j=1}^{k}A_{j}\otimes\overline{A_{j}}\right)  \operatorname{vec}\left(
\rho\right)  ,
\]
and therefore
\[
K\left(  \Phi\right)  :=\sum_{j=1}^{k}A_{j}\otimes\overline{A_{j}}.
\]
(Here $\overline{A_{j}}$ represents the entry-wise complex conjugate of
$A_{j}$.)

In the section that follows, we will make use of the following simple way to
calculate the natural matrix representation of a quantum operation that is
specified by a quantum circuit. Suppose that $\mathcal{R}$ is an $r$-qubit
system, $\mathcal{S}$ is an $s$-qubit system, and $U$ is a unitary operation
on $r+s$ qubits. Then for the quantum operation $\Phi$ defined as
\[
\Phi(\rho)=\operatorname*{Tr}\nolimits_{\mathcal{S}}\left[  U\left(
\rho\otimes\left(  |0\rangle\langle0|\right)  ^{\otimes s}\right)  U^{\dagger
}\right]  ,
\]
we have
\begin{equation}
K(\Phi)=M_{1}\left(  U\otimes\overline{U}\right)  M_{0} \label{eq:K(Phi)}%
\end{equation}
for
\[
M_{1}=\sum_{y\in\{0,1\}^{s}}I\otimes\left\langle y\right\vert \otimes
I\otimes\left\langle y\right\vert \quad\quad\text{and}\quad\quad
M_{0}=I\otimes\left\vert 0\right\rangle ^{\otimes s}\otimes I\otimes\left\vert
0\right\rangle ^{\otimes s}.
\]
(In both cases, each identity matrix $I$ acts on $\mathcal{R}$, or
equivalently is the $2^{r}\times2^{r}$ identity matrix.)

\subsection{Space-Bounded and Depth-Bounded Computations\label{SPACEBOUND}}

When we speak of a family $\{C_{n}\,:\,n\in\mathbb{N}\}$ of Boolean circuits,
we assume that each $C_{n}$ is an acyclic circuit, composed of AND, OR, NOT,
and constant-sized fanout gates, with $n$ input bits and an arbitrary number
of output bits. Such a family computes a function $f:\{0,1\}^{\ast}%
\rightarrow\{0,1\}^{\ast}$ if, for each $n\in\mathbb{N}$ and string
$x\in\{0,1\}^{n}$, the circuit $C_{n}$ outputs $f(x)$ when given input $x$.
\ The \textit{depth} of a Boolean circuit $C$ is the length of the longest
path in $C$ from an input bit to an output bit. \ The \textit{size} of $C$ is
the sum of the number of input bits, output bits, and gates.

For a given function $s:\mathbb{N}\rightarrow\mathbb{N}$, we say that a
Boolean circuit family $\{C_{n}\,:\,n\in\mathbb{N}\}$ is \textit{space
$O(s)$-uniform} if there exists a deterministic Turing machine $M$ that runs
in space $O(s)$, and that outputs a description of $C_{n}$ on input $1^{n}$
for each $n\in\mathbb{N}$. As is usual when discussing space-bounded
computation, a deterministic Turing machine is assumed to be equipped with a
read-only input tape that does not contribute to the space it uses, so it is
meaningful to consider sublinear space bounds. \ Given a space $O(s)$-uniform
family $\{C_{n}\,:\,n\in\mathbb{N}\}$, the size of $C_{n}$\ can be at most
$2^{O(s(n))}$.

We say a function $f:\{0,1\}^{\ast}\rightarrow\{0,1\}^{\ast}$ is in the class
$\mathsf{NC}\left(  s\right)  $ if there exists a space $O(s)$-uniform family
of Boolean circuits $\{C_{n}\,:\,n\in\mathbb{N}\}$ that computes $f$, and
where the depth of $C_{n}$ is at most$\ s\left(  n\right)  ^{O\left(
1\right)  }$.\footnote{$\mathsf{NC}$ stands for \textquotedblleft Nick's
Class\textquotedblright; the term is historical. \ Also, what we call
$\mathsf{NC}\left(  s\right)  $\ is called $\mathsf{NC}\left(  2^{s}\right)
$\ by Borodin, Cook, and Pippenger \cite{bcp}.} \ Also, a language $L$ is in
$\mathsf{NC}(s)$ if its characteristic function is in $\mathsf{NC}(s)$. \ We
write $\mathsf{NC}$ for $\mathsf{NC}(\log n)$, and $\mathsf{NC}\left(
\operatorname*{poly}\right)  $ for the union of $\mathsf{NC}(n^{c})$ over all
constants $c$. Borodin \cite{borodin} proved that if $s$ satisfies
$s(n)=\Omega\left(  \log n\right)  $, then every function in $\mathsf{NC}(s)$
is computable by a deterministic Turing machine in space $s(n)^{O(1)}$. \ It
follows that $\mathsf{NC}\left(  \operatorname*{poly}\right)  \subseteq
\mathsf{PSPACE}$. \ (The reverse containment also holds, so in fact we have
$\mathsf{NC}\left(  \operatorname*{poly}\right)  =\mathsf{PSPACE}$.)

It is clear that if $f\in\mathsf{NC}\left(  \operatorname*{poly}\right)  $ and
$g\in\mathsf{NC}$, then their composition $g\circ f$ is in $\mathsf{NC}\left(
\operatorname*{poly}\right)  $, since we can create a circuit for $g\circ f$
by composing the circuits for $f$ and $g$ in the obvious way.

Many functions of matrices are known to be computable in $\mathsf{NC}$.
\ These include sums and products of matrices, inverses, and the trace,
determinant, and characteristic polynomial, all over a wide range of fields
for which computations can be efficiently performed. \ (See von zur Gathen
\cite{vonzur}.) \ In particular, we will rely on a fact that follows from a
result of Borodin, Cook, and Pippenger \cite[Section 4]{bcp}:

\begin{theorem}
[\cite{bcp}]\label{bcpthm}The determinant of an $n\times n$ matrix whose
entries are rational functions in an indeterminate $z$ can be computed in
$\mathsf{NC}$.
\end{theorem}

\subsection{Projecting Onto Fixed Points\label{FIXEDPOINT}}

In this subsection, we prove a general theorem about efficient construction of
quantum operations that project onto the fixed-points of other quantum
operations. \ This theorem is the technical core of our $\mathsf{BQP}%
_{\mathsf{CTC}}\subseteq\mathsf{PSPACE}$\ result, but it might be of
independent interest as well.

\begin{theorem}
\label{crux}Suppose that $\Phi:\mathbb{C}^{N\times N}\rightarrow
\mathbb{C}^{N\times N}$ is a given quantum operation acting on an
$N$-dimensional system, meaning that it is a completely positive and
trace-preserving linear map. We will prove that there exists another quantum
operation
\[
\Lambda:\mathbb{C}^{N\times N}\rightarrow\mathbb{C}^{N\times N}%
\]
that satisfies three properties:

\begin{enumerate}
\item[(1)] For every density matrix $\sigma\in\mathbb{C}^{N\times N}$, it
holds that $\rho=\Lambda(\sigma)$ is a fixed-point of $\Phi$.

\item[(2)] Every density matrix $\rho$ that is a fixed-point of $\Phi$ is also
a fixed point of $\Lambda$.

\item[(3)] $\Lambda$ can be computed from $\Phi$ in $\mathsf{NC}$.
\end{enumerate}
\end{theorem}

In essence, $\Lambda$ is a (non-orthogonal) projection onto fixed-points of
$\Phi$, so if we want a fixed-point of $\Phi$ it suffices to compute
$\Lambda(\sigma)$ for any density matrix $\sigma$, and moreover every
fixed-point $\rho$ of $\Phi$ arises in this way from some density matrix
$\sigma$ (which always includes the choice $\sigma=\rho$).

\begin{proof}
The operation $\Lambda$ is defined as follows. First, for each real number
$z\in(0,1)$, we define a superoperator $\Lambda_{z}:\mathbb{C}^{N\times
N}\rightarrow\mathbb{C}^{N\times N}$ as
\[
\Lambda_{z}=z\sum_{k=0}^{\infty}(1-z)^{k}\Phi^{k}.
\]
Here $\Phi^{k}$ represents the $k$-fold composition of $\Phi$ and $\Phi^{0}$
is the identity operation. Each $\Phi^{k}$ is obviously completely positive
and trace-preserving. Given that $z(1-z)^{k}\in(0,1)$ for each choice of
$z\in(0,1)$ and $k\geq0$, and that $\sum_{k=0}^{\infty}z(1-z)^{k}=1$\ for
every $z\in(0,1)$, we have that $\Lambda_{z}$ is a convex combination of
completely positive and trace-preserving maps. Thus, $\Lambda_{z}$ is
completely positive and trace-preserving as well. Finally, we take
\[
\Lambda=\lim_{z\downarrow0}\Lambda_{z}.
\]
We must of course prove that this limit exists---and in the process, we will
prove that $\Lambda$ can be produced from $\Phi$ by an $\mathsf{NC}$
computation, which is an important ingredient of our simulation of
$\mathsf{BQP}_{\mathsf{CTC}}$ in $\mathsf{PSPACE}$. \ Once this is done, the
required properties of $\Lambda$ will be easily verified.

We assume that $\Phi$ is represented by the $N^{2}\times N^{2}$ complex matrix
$M=K(\Phi)$ as discussed in Section \ref{MATREP}. \ Since $\Phi$ is a quantum
operation, every eigenvalue of $M$ lies within the unit circle.\footnote{This
fact is proved in \cite{td:equilib}. An alternate proof follows from the fact
that $\left\Vert \Phi\right\Vert _{\diamond}=1$ and that the diamond norm is
submultiplicative (see~Theorem 5.6.9 of \cite{horn}).} \ It follows that the
matrix $I-(1-z)M$ is invertible for every real $z\in(0,1)$, and moreover there
is a convergent series for its inverse:
\begin{equation}
\left(  I-(1-z)M\right)  ^{-1}=I+(1-z)M+(1-z)^{2}M^{2}+\cdots\label{eq:series}%
\end{equation}
Now, for every $z\in(0,1)$ we define an $N^{2}\times N^{2}$ matrix $R_{z}$ as
follows:
\[
R_{z}:=z\,(I-(1-z)M)^{-1}.
\]
We note that $R_{z}=K(\Lambda_{z})$ for $\Lambda_{z}$ as defined above---and
as each $\Lambda_{z}$ is completely positive and trace-preserving, each entry
of $R_{z}$ must be bounded in absolute value by $1$.

Next, by Cramer's rule, we have
\begin{equation}
z\,\left(  I-(1-z)M\right)  ^{-1}\left[  i,j\right]  =(-1)^{i+j}%
\,\frac{z\,\det(\left(  I-(1-z)M\right)  _{j,i})}{\det\left(  I-(1-z)M\right)
}, \label{eq:cramer}%
\end{equation}
where $\left(  I-(1-z)M\right)  _{j,i}$ denotes the $(N^{2}-1)\times(N^{2}-1)$
matrix obtained by removing the $j^{th}$ row and $i^{th}$ column from
$I-(1-z)M$. \ It follows that each entry of $R_{z}$ is given by a rational
function in the variable $z$ having degree at most~$N^{2}$. \ As the entries
of $R_{z}$ are rational functions that are bounded for all $z\in(0,1)$, we
have that the limit $\lim_{z\downarrow0}R_{z}$ exists. \ Define
\[
R:=\lim_{z\downarrow0}R_{z},
\]
and note that $R=K(\Lambda)$. We have therefore proved that the limit
$\Lambda=\lim_{z\downarrow0}\Lambda_{z}$ exists as claimed.

The fact that $R$ can be computed from $M$ in \textsf{NC} follows from the
above discussion, together with Theorem \ref{bcpthm}. \ In particular,
equation \eqref{eq:cramer} above expresses the entries of $R_{z}$ as ratios of
polynomials of degree at most $N^{2}$ in $z$ having coefficients with rational
real and imaginary parts. \ It remains to compute the limit, which is also
done symbolically for the real and imaginary parts of each entry. \ To compute%
\[
\lim_{z\downarrow0}\frac{f(z)}{g(z)}%
\]
for polynomials $f\left(  z\right)  =\sum_{i}c_{i}z^{i}$\ and $g\left(
z\right)  =\sum_{i}d_{i}z^{i}$, we perform a binary search on the coefficients
of $g$ to find the smallest $k$ for which $d_{k}\neq0$, and then output the
ratio $c_{k}/d_{k}$. \ Each of the required computations can be done in
\textsf{NC}, and can be applied in parallel for each entry of $R$ to allow $R$
to be computed from $M$ in \textsf{NC}.

Finally, we verify the required properties of $\Lambda$. It is clear that
every fixed-point $\rho$ of $\Phi$ is also a fixed-point of $\Lambda$, since%
\[
\Lambda_{z}(\rho)=z\sum_{k=0}^{\infty}(1-z)^{k}\Phi^{k}(\rho)=z\sum
_{k=0}^{\infty}(1-z)^{k}\rho=\rho,
\]
and therefore $\Lambda(\rho)=\lim_{z\downarrow0}\Lambda_{z}(\rho)=\rho$. \ To
prove that $\rho=\Lambda(\sigma)$ is a fixed-point of $\Phi$ for every density
matrix $\sigma$, it suffices to prove $\Phi\Lambda=\Lambda$. For each
$z\in(0,1)$ we have
\[
\Phi\Lambda_{z}=z\,\sum_{k=0}^{\infty}(1-z)^{k}\Phi^{k+1}=\frac{z}{1-z}%
\,\sum_{k=1}^{\infty}(1-z)^{k}\Phi^{k}=\frac{1}{1-z}\Lambda_{z}-\frac{z}%
{1-z}I,
\]
and therefore
\[
\Phi\Lambda=\lim_{z\downarrow0}\Phi\Lambda_{z}=\lim_{z\downarrow0}\left(
\frac{1}{1-z}\Lambda_{z}-\frac{z}{1-z}I\right)  =\Lambda
\]
as claimed.
\end{proof}

\subsection{Proof of Containment\label{THEPROOF}}

We can now complete the proof that quantum computers with CTCs are simulable
in $\mathsf{PSPACE}$.

\begin{theorem}
\label{bqpctcthm} $\mathsf{BQP}_{\mathsf{CTC}}\subseteq\mathsf{PSPACE}$.
\end{theorem}

\begin{proof}
Let $L\in\mathsf{BQP}_{\mathsf{CTC}}$ be given, and assume that $\mathcal{A}$
is a quantum CTC algorithm for $L$. \ As discussed in Section~\ref{SPACEBOUND}%
, it suffices to prove $L\in\mathsf{NC}\left(  \operatorname*{poly}\right)  $.

Assume for simplicity that an input $x$ of length $n$ has been fixed. \ Let
$Q$ be the unitary quantum circuit that is output by $\mathcal{A}$ on input
$x$; then as in the definition of $\mathsf{BQP}_{\mathsf{CTC}}$, define a
quantum operation
\[
\Phi\left(  \rho\right)  :=\operatorname*{Tr}\nolimits_{\mathcal{R}_{CR}%
}\left(  Q\left(  \rho\otimes\left(  |0\rangle\langle0|\right)  ^{\otimes
r(n)}\right)  Q^{\dagger}\right)  .
\]
Our goal will be to compute the probability
\begin{equation}
\operatorname*{Pr}\left[  \mathcal{M}\left(  Q\left(  \rho\otimes\left(
|0\rangle\langle0|\right)  ^{\otimes r(n)}\right)  Q^{\dagger}\right)
=1\right]  \label{eq:probability-to-accept}%
\end{equation}
for some arbitrary fixed-point $\rho$ of $\Phi$. \ This value can then be
compared to $1/2$ to decide whether to accept or reject. \ This computation
will be performed in a uniform manner, in $\mathsf{NC}\left(
\operatorname*{poly}\right)  $, therefore establishing that $L\in
\mathsf{NC}\left(  \operatorname*{poly}\right)  $.

The first step is to compute the matrix representation $M=K(\Phi)$ of the
operation $\Phi$. This can be done by a polynomial-space uniform family of
Boolean circuits with exponential size and polynomial depth, since $M$ is
expressible as in equation \eqref{eq:K(Phi)}, and $Q$ is expressible as a
product of a polynomial number of exponential-size matrices determined by the
gates of $Q$.

Next we compute the matrix representation $R=K(\Lambda)$, where $\Lambda$ is
the quantum operation that projects onto fixed-points of $\Phi$ discussed in
Section~\ref{FIXEDPOINT}. We have argued that $R$ can be computed from $M$ in
$\mathsf{NC}$, and therefore by composing this computation with the
$\mathsf{NC}\left(  \operatorname*{poly}\right)  $ computation of $M$, we have
that $R$ can be computed in $\mathsf{NC}\left(  \operatorname*{poly}\right)  $.

Finally, we compute a fixed-point $\rho$ of $\Phi$ using $R$ along with an
arbitrary choice of a density matrix input for $\Lambda$. For instance, we may
take $\rho=\Lambda\left(  \left(  \left\vert 0\right\rangle \left\langle
0\right\vert \right)  ^{\otimes q(n)}\right)  $, so that $\operatorname*{vec}%
(\rho)=R\,\left\vert 0\right\rangle ^{\otimes2q(n)}$. \ The probability
\eqref{eq:probability-to-accept} can then be evaluated in $\mathsf{NC}\left(
\operatorname*{poly}\right)  $ by performing matrix-vector multiplication.
\end{proof}

\section{Dealing With Error\label{ERROR}}

Recall that, in defining the class $\mathsf{BQP}_{\mathsf{CTC}}$, we required
the quantum circuits to involve amplitudes with rational real and imaginary
parts. \ However, while this assumption is mathematically convenient, it is
also \textquotedblleft unphysical.\textquotedblright\ \ Even in a CTC
universe, quantum operations can presumably only be implemented to finite precision.

The trouble is that, in a CTC universe, two quantum operations that are
arbitrarily close can produce detectably different outcomes. \ As an example,
consider the stochastic matrices%
\[
\left(
\begin{array}
[c]{cc}%
1 & \varepsilon\\
0 & 1-\varepsilon
\end{array}
\right)  ,\left(
\begin{array}
[c]{cc}%
1-\varepsilon & 0\\
\varepsilon & 1
\end{array}
\right)  .
\]
As $\varepsilon\rightarrow0$, these matrices become arbitrarily close to each
other and to the identity. \ Yet their fixed-points remain disjoint: the first
has a unique fixed-point of $\left(  1,0\right)  ^{T}$, while the second has a
unique fixed-point of $\left(  0,1\right)  ^{T}$. \ Hence, were an algorithm
to apply one of these matrices inside a CTC, an arbitrarily small error could
completely change the outcome of the computation.

We will show that, while this \textquotedblleft pathological\textquotedblright%
\ situation can arise in principle, it does not arise in our simulation of
$\mathsf{PSPACE}$\ by a CTC computer in Lemma \ref{inpctc}.

Call $\rho$\ an $\varepsilon$\textit{-fixed-point} of $\Phi$ if $\left\Vert
\rho-\Phi\left(  \rho\right)  \right\Vert _{\operatorname*{tr}}\leq
\varepsilon$. \ 

\begin{proposition}
\label{epsfp}Suppose $\rho$\ is a fixed-point of $\Phi$ and $\left\Vert
\Phi-\Phi^{\prime}\right\Vert _{\diamond}\leq\varepsilon$. \ Then $\rho$\ is
an $\varepsilon$-fixed-point\ of $\Phi^{\prime}$.
\end{proposition}

\begin{proof}
Since $\rho=\Phi\left(  \rho\right)  $, we have $\left\Vert \rho-\Phi^{\prime
}\left(  \rho\right)  \right\Vert _{\operatorname*{tr}}=\left\Vert \Phi\left(
\rho\right)  -\Phi^{\prime}\left(  \rho\right)  \right\Vert
_{\operatorname*{tr}}\leq\varepsilon$.
\end{proof}

\begin{lemma}
\label{cycle}Let $\Phi$ be a classical operation mapping a finite set
$\mathcal{B}$\ to itself, and let $\rho$\ be an $\varepsilon$-fixed-point of
$\Phi$. \ Then $\left\Vert \rho-\sigma\right\Vert _{\operatorname*{tr}}%
\leq2\left\vert \mathcal{B}\right\vert \varepsilon$ for some state $\sigma
$\ supported only on the cycles of $\Phi$.
\end{lemma}

\begin{proof}
We prove the contrapositive. \ Let $\mathcal{C}$\ be the union of all cycles
of $\Phi$, and let $\overline{\mathcal{C}}=\mathcal{B}\setminus\mathcal{C}$.
\ Also, for each element $x\in\mathcal{B}$, let $p_{x}=\left\langle
x\right\vert \rho\left\vert x\right\rangle $. \ Suppose $\rho$\ is not
$\delta$-close to any state supported only on $\mathcal{C}$, where
$\delta=2\left\vert \mathcal{B}\right\vert \varepsilon$. \ Then $\sum
_{x\in\overline{\mathcal{C}}}p_{x}>\delta$. \ Hence%
\begin{align*}
\left\Vert \rho-\Phi\left(  \rho\right)  \right\Vert _{\operatorname*{tr}}  &
\geq\frac{1}{2}\sum_{x\in\overline{\mathcal{C}}}\left\vert p_{x}%
-p_{\Phi\left(  x\right)  }\right\vert \\
&  \geq\frac{1}{2}\max_{x\in\overline{\mathcal{C}}}p_{x}\\
&  >\frac{1}{2}\cdot\frac{\delta}{\left\vert \overline{\mathcal{C}}\right\vert
}\\
&  \geq\varepsilon.
\end{align*}

\end{proof}

Let $C^{\prime}$\ be the circuit from Lemma \ref{inpctc}\ that maps
$\mathcal{R}_{CTC}$\ to itself. \ As part of the proof of Lemma \ref{inpctc},
we showed that the graph of $C^{\prime}:\left\{  0,1\right\}  ^{p\left(
n\right)  }\rightarrow\left\{  0,1\right\}  ^{p\left(  n\right)  }$\ has a
unique cycle $L$, in which every configuration leads to the desired output.
\ Now let $C^{\prime\prime}$\ be a corrupted version of $C^{\prime}$ that
satisfies $\left\Vert C^{\prime}-C^{\prime\prime}\right\Vert _{\diamond}%
\leq\varepsilon$, and let $\rho$\ be any fixed-point of $C^{\prime\prime}$.
\ Then $\rho$\ is an $\varepsilon$-fixed-point of $C^{\prime}$\ by Proposition
\ref{epsfp}. \ By Lemma \ref{cycle}, this in turn means that $\left\Vert
\rho-\sigma\right\Vert _{\operatorname*{tr}}\leq2^{p\left(  n\right)
+1}\varepsilon$ for some state $\sigma$\ supported only on $L$. \ So provided
$\varepsilon\ll2^{-p\left(  n\right)  -1}$, a CTC algorithm that uses
$C^{\prime\prime}$\ in place of $C^{\prime}$\ will still produce the correct
answer with high probability.

Moreover, as pointed out by Bacon \cite{bacon}, even if every gate in our
quantum circuit is subject to constant error, we can still use standard
results from the theory of quantum error-correction \cite{ab}\ to ensure that
$\left\Vert C^{\prime}-C^{\prime\prime}\right\Vert _{\diamond}$\ is
exponentially small, where $C^{\prime}$\ and $C^{\prime\prime}$\ are the
quantum circuits acting on the \textit{logical} (encoded) qubits.

One might also ask whether our proof of $\mathsf{BQP}_{\mathsf{CTC}}%
\subseteq\mathsf{PSPACE}$\ in Section \ref{QUANTUM}\ is affected by precision
issues. \ However, the key point there is that, since the amplitudes are
assumed to be complex rational numbers, we are able to use the algorithm of
Borodin, Cook, and Pippenger \cite{bcp}\ to\ compute a fixed-point
\textit{symbolically} rather than just numerically.

\section{Discussion and Open Problems\label{DISC}}

\subsection{CTCs in Other Computational Models\label{OTHERMODEL}}

In the proof that $\mathsf{PSPACE}\subseteq\mathsf{P}_{\mathsf{CTC}}$, we did
not actually need the full strength of polynomial-time computation inside the
CTC: rather, all we needed was the ability to update the configuration of a
$\mathsf{PSPACE}$\ machine and increment a counter. \ Thus, our proof also
shows (for example) that $\mathsf{PSPACE}=\mathsf{AC}_{\mathsf{CTC}}^{0}$,
where $\mathsf{AC}^{0}$\ denotes the class of problems solvable by
constant-depth, polynomial-size circuits consisting of AND, OR, and NOT gates,
and $\mathsf{AC}_{\mathsf{CTC}}^{0}$\ is defined the same way as
$\mathsf{P}_{\mathsf{CTC}}$\ but with $\mathsf{AC}^{0}$\ circuits instead of
arbitrary polynomial-size circuits.

In the other direction, we could also define $\mathsf{PSPACE}_{\mathsf{CTC}}$
the same way as $\mathsf{P}_{\mathsf{CTC}}$, but with $\mathsf{PSPACE}%
$\ machines in place of polynomial-size circuits. \ Then it is evident that
our proof generalizes to show $\mathsf{PSPACE}_{\mathsf{CTC}}=\mathsf{PSPACE}$.

It would be extremely interesting to study the consequences of Deutsch's
causal consistency assumption in other settings besides polynomial-time
computation: for example, communication complexity, branching programs, and
finite automata.

\subsection{Narrow CTCs\label{ONEBIT}}

What is the power of classical CTCs with a single bit, or of quantum CTCs with
a single qubit (as studied by Bacon \cite{bacon})? \ Let $\mathsf{P}%
_{\mathsf{CTC1}}$, $\mathsf{BPP}_{\mathsf{CTC1}}$, and $\mathsf{BQP}%
_{\mathsf{CTC1}}$\ be the corresponding deterministic, randomized, and quantum
complexity classes. \ Then it is not hard to show that $\mathsf{NP\cap
coNP}\subseteq\mathsf{BPP}_{\mathsf{CTC1}}$: that is, a single use of a
one-bit CTC is enough to solve all problems in the class $\mathsf{NP\cap
coNP}$. \ For we can guess a random string $w\in\left\{  0,1\right\}
^{p\left(  n\right)  }$, then set the CTC bit $b$ to $1$ if $w$ is a
yes-witness or to $0$ if $w$ is a no-witness, and leave $b$ unchanged if
$w$\ is neither. \ If there exists a yes-witness but not a no-witness, then
the only fixed-point of the induced stochastic evolution is $b=1$, while if
there exists a no-witness but not a yes-witness, then the only fixed-point is
$b=0$. \ Indeed, a simple extension of this idea yields $\mathsf{NP}%
\subseteq\mathsf{BPP}_{\mathsf{CTC1}}$: we set $b=1$\ if a yes-witness
$w\in\left\{  0,1\right\}  ^{p\left(  n\right)  }$ was guessed, and set $b=0$
with some tiny probability $\varepsilon\ll2^{-p\left(  n\right)  }%
$\ independent of the witness. \ Again, the unique fixed-point of the induced
stochastic map will fix $b=1$\ with overwhelming probability if there exists a
yes-witness, or $b=0$\ with certainty if not. \ Fully understanding the power
of \textquotedblleft bounded-width CTCs\textquotedblright\ remains a problem
for the future.

\subsection{CTC Computers With Advice\label{ADVICE}}

Let $\mathsf{BPP}_{\mathsf{CTC}}\mathsf{/rpoly}$\ be defined the same way as
$\mathsf{BPP}_{\mathsf{CTC}}$,\ except that instead of being initialized to
$0^{q\left(  n\right)  }$, the chronology-respecting register $\mathcal{R}%
_{CR}$ is initialized to a probability distribution $\mathcal{D}_{n}$\ which
depends only on the input length $n$, but can otherwise be chosen arbitrarily
to help the CTC algorithm. \ Then we claim that$\ \mathsf{BPP}_{\mathsf{CTC}%
}\mathsf{/rpoly}=\mathsf{ALL}$: in other words, $\mathsf{BPP}_{\mathsf{CTC}%
}\mathsf{/rpoly}$\ contains \textit{every} computational problem! \ To see
this, let $\mathcal{R}_{CR}$\ be initialized to the uniform distribution over
all ordered pairs $\left\langle z,f\left(  z\right)  \right\rangle $, where
$z$\ is an $n$-bit input and $f\left(  x\right)  \in\left\{  0,1\right\}
$\ encodes whether $x\in L$. \ Also, let the CTC register $\mathcal{R}_{CTC}$
contain a single bit $b$. \ Then given an input $x$, our circuit $C$ acts on
$b$ as follows: if $z=x$\ then $C$ sets $b=f\left(  x\right)  $; otherwise $C$
leaves $b$ unchanged. \ It is easy to see that the unique fixed-point of the
induced stochastic map on $\mathcal{R}_{CTC}$\ fixes $b=f\left(  x\right)  $
with certainty.

While it demonstrates that CTCs combined with randomized advice yield
staggering computational power, this result is not \textit{quite} as
surprising as it seems: for it was previously shown by Aaronson
\cite{aar:qmaqpoly} that $\mathsf{PP/rpoly}=\mathsf{ALL}$, and by Raz
\cite{raz:all}\ that $\mathsf{IP}\left(  2\right)  \mathsf{/rpoly}%
=\mathsf{ALL}$. \ In other words, randomized advice has a well-known tendency
to yield unlimited computational power when combined with certain other resources.

\section{Acknowledgments}

We thank Lance Fortnow for suggesting the proof of Lemma \ref{inpctc}, and
Dave Bacon for helpful discussions.

\bibliographystyle{plain}
\bibliography{thesis}

\end{document}